# PLUTO'S LIGHT CURVE IN 1933-1934

**Bradley E. Schaefer** (Department of Physics and Astronomy, Louisiana State University, Baton Rouge Louisiana 70803 ),
**Marc W. Buie** (Lowell Observatory, Flagstaff Arizona 86001),
**Luke Timothy Smith** (Department of Physics and Astronomy, Louisiana State University, Baton Rouge Louisiana 70803 )

## ABSTRACT

The Pluto-Charon system has complex photometric variations on all time scales; due to rotational modulations of dark markings across the surface, the changing orientation of the system as viewed from Earth, occultations and eclipses between Pluto and Charon, as well as the sublimation and condensation of frosts on the surface. The earliest useable light curve for Pluto is from 1953-1955 when Pluto was 35 AU from the Sun. Earlier data on Pluto has the potential to reveal properties of the surface at a greater heliocentric distance with nearly identical illumination and viewing geometry. We are reporting on a new accurate photographic light curve of Pluto for 1933-1934 when the heliocentric distance was 40AU. We used 43 B-band and V-band images of Pluto on 32 plates taken on 15 nights from 19 March 1933 to 10 March 1934. Most of these plates were taken with the Mount Wilson 60" and 100" telescopes, but 7 of the plates (now at the Harvard College Observatory) were taken with the 12" and 16" Metcalf doublets at Oak Ridge. The plates were measured with an iris diaphragm photometer, which has an average one-sigma photometric error on these plates of 0.08 mag as measured by the repeatability of constant comparison stars. The modern B and V magnitudes for the comparison stars were measured with the Lowell Observatory Hall 1.1-m telescope. The magnitudes in the plate's photographic system were converted to the Johnson B- and V-system after correction with color terms, even though they are small in size. We find that the average B-band mean opposition magnitude of Pluto in 1933-1934 was 15.73 ± 0.01, and we see a roughly sinusoidal modulation on the rotational period (6.38 days) with a peak-to-peak amplitude of 0.11 ± 0.03 mag. With this, we show that Pluto darkened by 5% from 1933-1934 to 1953-1955. This darkening from 1933-1934 to 1953-1955 cannot be due to changing viewing geometry (as both epochs had identical sub-Earth latitudes), so our observations must record a real albedo change over the southern hemisphere. The later darkening trend from 1954 to the 1980's has been explained by changing viewing geometry (as more of the darker northern hemisphere comes into view). Thus, we now have strong evidence for albedo changes on the surface of Pluto, and these are most easily explained by the systematic sublimation of frosts from the sunward pole that led to a drop in the mean surface albedo.

## 1. INTRODUCTION

The discovery of Pluto was a major astronomical event of the 20th century. On April 4, 1929, as part of a dedicated search at Lowell Observatory for trans-Neptunian planets, Clyde Tombaugh took the first of his many photographic plates, and implemented a calculated, systematic method using a blink comparator. He discovered Pluto on February 18, 1930 using exposures taken on January 23, 1930 and January 29, 1930. The announcement led to a flurry of deep photographs aimed at measuring Pluto's position so as to allow for an accurate orbit. As part of this effort, prediscovery images were sought, with 14 being reported from the years 1914 to 1927 (Bower 1931).

Almost half a century later, in 1978, James W. Christy noticed southern and northern elongations on Pluto images, with this being the discovery of Pluto's moon Charon (Christy and Harrington 1978). Charon's orbit allowed a determination of Pluto's mass, and this was found to be so small as to invalidate the success of Percival Lowell's predictions. Indeed, later work has shown that the anomalies in Uranus' and Neptune's orbits that were the basis of these predictions have been removed with modern ephemerides (Standish 1993). Charon's orbital period equaled the previously discovered 6.38 day photometric modulation (Walker and Hardie 1955), so that it was apparent that Pluto and Charon were in synchronous rotation with albedo markings causing the modulation. Two more moons of Pluto (Nix and Hydra) were discovered recently with the *Hubble Space Telescope* (Weaver et al. 2006).

Pluto was widely observed soon after its discovery, with an early light curve being given by Walter Baade using plates from the Mount Wilson 60" and 100" telescopes (Baade 1934). The early light curves were all obviously far wrong, so these magnitudes have always been largely disregarded (e.g., Marcialis 1997). The primary observational task in this paper will be to derive reliable light curves from the plates taken by Baade. The first photoelectric light curve (with much more reliable magnitudes) was from 1953-1955 (Walker and Hardie 1955), with later light curves in 1964 (Hardie 1965), 1971-1973 (Andersson and Fix 1973), and 1980-1983 (Tholen and Tedesco 1994). Starting in the 1980's, with the Pluto-Charon mutual events spurring accurate photometry ( see Binzel and Hubbard 1997 for a review), the system brightness was extensively monitored, resulting in a map of light and dark spots across the surface (see Buie, Young and Binzel 1997 for a review). In all, the photometric history of Pluto is well observed from 1953 to present. The goal of this paper is to extend Pluto's photometric history from the 1953-1955 back to 1933, with this two decade interval including substantial sublimation of surface frosts.

## 2. PHOTOMETRY

The old photographic plates are still in pristine shape. Modern CCD images have a substantially better photometric precision than does photographic photometry (typically 0.01 mag versus 0.1 mag). As such, the current generation of astronomers tend to ignore any old photographic plates. Indeed, few working astronomers have now ever handled astronomical plates and even fewer have any idea of how to analyze the plates so as to measure magnitudes. This ignorance is a loss to our community for the many front-line questions for which the time dependence of a phenomenon or past events is critical. In the case of Pluto, the old plates can provide pre-modern light curves as the only information about frost migration as Pluto approached the Sun.

A substantial problem with using photometry from old papers is that their adopted magnitudes for comparison stars have large systematic errors, with the old values being too bright by 0.5-1.0 mag. For example with Baade's light curve of Pluto from 1933-1934 (Baade 1934), his comparison star magnitudes were derived by photographic transfer from nearby Selected Areas, and these magnitudes are too bright by 0.5-1.0 mag (Sandage 2001). This systematic problem with comparison star magnitudes has invalidated many of the results from old papers. The general solution is to remeasure the comparison stars in modern times, and possibly to remeasure the original plates themselves. We have long experience with modern analysis of old plates as applied to current front-line astrophysics (Schaefer 1990; 1994; 1995; 1996; 2005; Schaefer and Fried 1991; Schaefer and Patterson 1983; Robinson, Clayton, and Schaefer 2006).

Our realization was that we could get old Pluto plates, derive a reliable modern light curve for the old plates, and that this is a valuable extension of the long-term photometric record of Pluto. To accomplish this, we had to locate the old plates, measure the instrumental magnitudes of Pluto and its comparison stars, measure the B and V magnitudes of these comparison stars, and derive the B and V magnitudes of Pluto. The only disadvantage from modern CCD results is that our median 1-sigma uncertainty is 0.08 mag (versus perhaps 0.01 mag for CCD measures). This uncertainty is fine for answering many questions about the early Pluto. And with many plates, the combined uncertainties can be substantially reduced. The result is a light curve from the early 1930's with all needed accuracy.

We searched for plates from the early 1930's, as this was a time when many plates were taken for astrometric purposes, and this is the earliest possible time to get a full light curve of Pluto. The Harvard plates were identified from an exhaustive search. These high quality images were taken with either the 12" Metcalfe refractor (the MA plate series with a plate scale of 97 "/mm) or the 16" Metcalfe refractor (the MC plate series with a plate scale of 98"/mm) at the Oak Ridge observatory in Massachusetts. The emulsion is only noted as "Hi-speed". The Mount Wilson plates were identified by John Grula (at Carnegie Observatories) based on details in Baade (1934). The Baade plates were taken with the 60" and 100" telescopes specifically to get Pluto magnitudes, with the time chosen being near quadrature so as to minimize Pluto's motion. The telescopes were always stopped down to effective apertures of 40" and 84" respectively ("to increase the field of definition"). The plates scales were 16 "/mm and 27 "/mm respectively. Several of Baade's plates were taken as multiple exposures with offsets between the individual exposures. Only six plates have the emulsion recorded on the envelope, and these are "E40" (for the October 26-27 1933 plates and for plate B296) and "Agfa Isodrom S1526" for plate B291. Four of Baade's images were taken with a V-band filter in front of the emulsion.

We identified and borrowed a total of 32 plates on 15 nights over a one-year interval from 1933 March 19 to 1934 March 10. With the intentional multiple exposures, this results in a total of 43 usable images of Pluto, of which 39 are in the B-band and 4 are in the V-band. The properties of these images are given in Table 1. The first column gives the plate number (plus a sub-exposure identification). The second column gives the UT date of the exposure. The third column gives the UT time of the middle of the exposure ($UT_{mid}$) with the time being known to the nearest minute. No corrections were made for light travel time. The fourth column gives the telescope aperture (in inches), with the 60" and 100" being on Mount Wilson and the 12" and 16" being at Oak Ridge. The fifth column gives the exposure time in minutes. The sixth column gives an evaluation of the image quality for Pluto and identifies which images are with the V-band filter. The seventh column gives the east longitude of the sub-Earth point on Pluto ($\lambda_{sub\oplus}$ in degrees), as appropriate for determining what albedo markings are facing Earth. We use the definition of north for Pluto based on the angular momentum vector and we are also using an east longitude scale to maintain a right-handed coordinate system as described in Buie et al. (1992). These longitudes are based on a rotational period of 6.387223 ± 0.000017 days (Tholen and Buie 1997) and also includes the effects of the Earth's motion. The next three columns give the Sun-Pluto distance (r in AU), the Earth-Pluto distance ($\Delta$ in AU), and the solar phase angle (the Sun-Pluto-Earth angle, $\alpha$, in degrees) all as taken from the JPL Horizons ephemeris program. Pluto is known to have very small opposition surge with a phase coefficient of 0.0372 ± 0.0016 mag $deg^{-1}$ (Tholen and Tedesco 1994). All but five of our plates were taken with $1.12<\alpha<1.41$. The last column gives a correction ($\Delta m$ in magnitudes) to the observed magnitudes that takes out the effects of distances and the phase curve. In particular, we give the correction to a mean opposition magnitude ($m_{opp}=m+\Delta m$) as $\Delta m=-5*Log_{10}[r\Delta/1520.7]-0.0372\alpha$, for a mean opposition (r=39.5 AU and $\Delta$=38.5 AU) with $r\Delta$=1520.7 $AU^2$.

The magnitudes of the comparison stars were measured on the nights of 2007 January 15, 17, and 18 with the CCD camera on the 1.1-m Hall telescope at Lowell Observatory. We also obtained absolute photometry of some comparison stars with the SMARTS 1.0-m and 1.3-m telescopes at Cerro Tololo InterAmerican Observatory. In addition, Arlo Landolt independently measured the magnitudes of six key comparison stars with the Blue Photometer on the 1.8-m telescope at Lowell Observatory.

Our photometry was of 17-27 comparison stars within each of the six fields with Pluto. These stars were chosen for proximity to Pluto positions, the lack of nearby stars, and for a range of magnitudes within roughly 1.5 mag of Pluto. All comparison stars were observed independently on 3 photometric nights, with the results being satisfactorily compared and then averaged together. All comparison stars had 23-38 separate good observations going into our final magnitudes. Images were taken in both the B-band and V-band, as this allows for color terms to be measured both in our photometry of the comparison stars with the modern CCD data as well as with the old photographs. Intermixed with the images of the Pluto fields, we also took many images of standard star fields (Landolt 1992) over a large range of airmass.

With our standard and comparison star data, we performed the usual analysis to derive the B and V magnitudes of the comparison stars. This included a calibration of the color terms so as to cast our photometry onto the scale of the Johnson magnitude systems. The statistical error from our CCD photometry had a median value of 0.007 mag in the B-band and 0.004 mag in the V-band. (The six faint stars with statistical errors of >0.05 mag in the B-band were excluded from the start of our analysis, although they would have been excluded later in our analysis as being far from the magnitude of Pluto.) Our systematic errors can be estimated by looking at the RMS scatter of the standard stars in their calibration plot as well as by the night-to-night repeatability. On this basis, we estimate our systematic uncertainty to be roughly 0.01 mag. In all, we have used modern CCD photometry to measure the magnitudes of many stars near Pluto in the B and V bands with an accuracy of roughly 0.01 mag, and these magnitudes also apply to the same stars back in the 1930's.

We present our comparison star magnitudes in Table 2. The first column is our internal star ID, with a field number followed by the star number. The second and third columns have the J2000 coordinates for each star. The fourth and fifth columns present the B and V magnitudes for the stars, with the 1-sigma measurement error being less than 0.01 mag in almost all cases.

The best way to measure the magnitudes of stars on photographic plates is to use an iris diaphragm photometer (IDP). (Detailed experimentation [e.g. Schaefer and Fried 1991; Schaefer 1982; 1995] and long experience has proven that the IDP measures are equally good as the various two-dimensional plate scanners.) The IDP is an analog means of getting an instrumental magnitude for a star that is similar to the aperture photometry (say, with APPHOT in IRAF) used for CCD images. With CCDs replacing photographs starting in the 1980's, IDPs have become scarce in the world. Fortunately, one of these is next to our offices at Louisiana State University, having been preserved by Arlo Landolt. This IDP was originally built by Askania.

We used the IDP to measure Pluto and many comparison stars (typically 17) on each plate. Many comparison stars, background points, and Pluto were measured multiple times. The whole process was repeated an average of twice per plate. This allows us to quantify our measurement errors.

For each star image measured, the IDP returns a dial reading that is a function of the aperture radius. We have calibrated the dial reading versus the physical radius of the diaphragm. Thus, we can determine the radius of the diaphragm (R) for every image. For every star, we also offset the diaphragm to a nearby blank patches of sky and take another reading from which we get the diaphragm radius for the background ($R_b$).

Photographic responses are nonlinear and this complicates analysis. Fortunately, Schaefer (1981) has proven that the quantity $R^2$-$R_b^2$ is linear with magnitude over a wide range, including that relevant for the Pluto images. In addition, this quantity is independent of any background variations (Schaefer 1979; 1981), although none of the Pluto plates show any background variation (with one exception as noted below) over the small regions with measured stars. Stars that are not saturated in the core (within roughly half a magnitude of the plate limit) will display small deviations from this linearity, while very bright stars (roughly >6 magnitudes brighter than the plate limit) will also have a nonlinear relation, with the correct functional forms derived in Schaefer (1981).

With the images of Pluto and its comparison stars being in the linear regime, a plot of $R^2$-$R_b^2$ versus magnitude will provide a calibration curve relating the instrumental measurement to the real magnitude. That is, for comparison stars within typically 0.8 mag of Pluto, a chi-square fit is made, with this then providing the relation to be applied to Pluto. However, the effective bandpass of the plate might have a color term in transforming to that of the standard Johnson B and V magnitude systems. To allow for this possibility, the chi-square fit is actually made for the model magnitude

$$m = m_0 + S\,(R^2\text{-}R_b^2) + C_S(B\text{-}V). \qquad (1)$$

Here, $m_0$ and S are fit parameters representing the best fit in the calibration curve, while $C_S$ is the linear correction for the deviations between the plate's bandpass and the Johnson photometric system. The B-V color for the comparison stars are known and the B-V color for Pluto is assumed to be 0.84 mag (Tholen & Buie 1997). This color might change by small amounts, but given the small size of the color terms (see next paragraph) any such changes will lead to errors that are much smaller than 0.01 mag. For each plate, 6-10 comparison stars were used in the chi-square fit to equation 1. To illustrate the calibration curves, we present three of them in Figure 1. We have selected our best, typical, and worst cases to present. The reasons for why some plates are better or worse than others is hard to know for certain, and they vary from plate-to-plate, but likely includes the usual differences in focus, extinction, sky brightness, and photographic development.

The color terms are expected to be small. This is because the plate emulsions used have a standard spectral sensitivity, and indeed the Harvard plates from the MA and MC series were used for the definition of the photographic magnitude system (which has a zero color term with respect to the Johnson B system). On a plate-by-plate basis, the color term can be evaluated by plotting and fitting $m\text{-}m_0\text{-}S\ (R^2\text{-}R_b^2)$ versus the B-V colors for the comparison stars. $C_S$ is found to be near zero to within the uncertainties for all the plates. Nevertheless, a more accurate measure of the color terms comes from averaging the values for all the plates from a single emulsion type. For this, the Harvard plates were found to have $C_S$=0.00; the B-band images from Mount Wilson have $C_S$=-0.07; and the V-band images from Mount Wilson have $C_S$=-0.06. If Pluto's B-V color changes from 0.84 mag to 0.78 mag (cf. Section 3), then the largest color term will lead to a systematic error of -0.07x(0.78-0.84)=0.004 mag, and this is greatly below the statistical uncertainty of even the all-combined average magnitude.

Our final derived magnitudes for Pluto are very insensitive to the analysis of the color terms. First, the color terms are near zero. Second, we find empirically that the derived magnitudes change by less than 0.01 mag for cases where we set the value of $C_S$ anywhere in the range from -0.3 to 0.3. Third, nearly all of our comparison stars have B-V values within 0.3 mag of the B-V of Pluto with a symmetric distribution. With this, the difference in the average B-V from that of Pluto is <0.1 mag and then the systematic error must be less than 0.01 mag.

The uncertainties in the derived magnitude for Pluto come from several sources. The dominant uncertainty arises from the normal noise in each image, with this being random fluctuations in the number of developed grains for a given flux. This grain noise is always present, even on the good quality plates that we are using, and it can be likened to ordinary noise from Poisson statistics in each CCD pixel. The uncertainties in the color terms are all small (see previous paragraph) and due to the small range of comparison star colors well centered on the color of Pluto. The measurement errors for the IDP are very small, as determined by the reproducibility from multiple measures. The uncertainty associated with the absolute calibration of the comparison stars is roughly 0.01 mag. The total uncertainties in the model magnitudes are represented by the RMS scatter of the observed comparison star magnitudes around their best fit model magnitudes. The 1-sigma uncertainty in the magnitude for Pluto is then set equal to this RMS scatter of the comparison stars around the best fit to equation 1.

How accurate is each magnitude? This can be quantified by looking at the scatter of the measures for the comparison stars around the calibration equation. We find that the typical scatter is 0.08 mag, although the range is from 0.04-0.18 mag (with most being between 0.06 and 0.12 mag). We find that half the plates give one-sigma uncertainties of 0.08 mag or better. This might be mildly surprising to people who are not experienced with photometry from top-quality old photographs. In principle, with 39

B-band images to beat down the random scatter, our *average* magnitude for Pluto will have a *statistical* error ($0.08/39^{0.5}$) approaching 0.01 mag.

With the collective *statistical* error being small, any *systematic* uncertainties will start to dominate. Given our above procedure (especially accounting for the color terms in both the magnitudes of the comparison stars and in the magnitudes from the plates), we cannot think of any significant systematic problems. For example, the trailing of Pluto's image is much smaller than the size of the star image (due in part to the images being taken near quadrature) and is completely negligible (cf. Schaefer 1981). The largest systematic error that we know about comes from the modern absolute calibration of our comparison stars, and this is accurate to roughly the 0.01 mag level. So we have no reason to suspect that the real errors are significantly worse that our quoted statistical errors. Indeed, the reduced chi-square for our phased light curve is near unity, and this implies that our quoted error bars are close to being correct.

In all, we have proven a method for deriving reliable modern magnitudes from the old plates with a typical 1-sigma uncertainty of 0.08 mag for a single plate. Our combined average magnitude will have an accuracy of roughly 0.01 mag.

## 3. LIGHT CURVE

With the procedures from the previous section, we have measured and derived the B-band magnitudes for 39 images and the V-band magnitudes for 4 images. Our magnitudes are presented in Table 3. The first column is the plate and image identification, while the second column gives the band (B or V) for the image. The third column is the Julian Date for the middle of the exposure with no corrections for light travel time. The fourth column gives our observed magnitude (m) with the 1-sigma uncertainty. The last column gives our mean opposition magnitudes corrected for the distance and phase of Pluto (with $m_{opp}=m+\Delta m$).

We have only one outlier point, and that is for plate M2068. Our measured magnitude is $m_{opp}=15.44 \pm 0.04$, while the other values for that phase of Pluto's rotation are $15.73 \pm 0.03$. This is a 5-sigma outlier and the magnitude can be tossed out on this basis alone. However, M2068 is the only plate that shows any variations in background density, with the reason being unknown and with Pluto being near the edge of a large change. On the basis of this apparent problem, the magnitude from M2068 will be disregarded further. Plates MA3601 and B150 are brighter and fainter from the phase average magnitudes respectively, but by less than the 2-sigma level with no independent cause for suspecting trouble, so these values are not considered as outliers.

The widest time span on any one night being 0.15 days. With this being small compared to the rotational period of Pluto (6.38 days), we can average together all the magnitudes from one night with no loss of resolution. Indeed, nightly averages can serve to substantially reduce the error bars. To this end, we have made weighted averages of each night's mean opposition B-band magnitudes ($B_{opp}$). These are presented in Table 4. The first two columns are the average $JD_{mid}$ and our nightly averaged $m_{opp}$ values in the B-band (i.e., $B_{opp}$). (The additional columns in the table give the model sub-Earth position and magnitudes as discussed in Section 5.) This is the primary observational result from this paper.

The weighted average B-band mean opposition magnitude of Pluto for 1933/1934 is $\langle B_{opp} \rangle=15.74 \pm 0.01$. With B-V=0.84 for Pluto, this translates to $V_{opp}=14.90 \pm 0.01$.

An alternative way to get the average B-band magnitude is to fit the phased light curve to a sinewave. This produces the average $\langle B_{opp} \rangle = 15.73 \pm 0.01$. From this fit, the full peak-to-peak amplitude is 0.11 ± 0.03 mag and the brightest rotational phase was a sub-Earth longitude of 260°.

Many of the nightly averages have similar phase and can be combined with no significant loss of resolution. We have made weighted averages for nights with similar phases, as listed in Table 5. Figure 2 presents a plot of the rotationally phased light curve. The best fit sinewave is superposed on the phase averages in Fig. 2. The modern light curve of Pluto requires higher order Fourier components, but our light curve does not have the accuracy to justify (for example by an F-test for adding terms) using any higher order terms.

We have four images in the V-band. This is not enough to get a useful independent light curve, but it is good enough to get the measured color to Pluto in 1933. For each of the four plates (with V-band magnitudes in Table 3) we can get the B-V magnitudes by subtracting off the nightly averaged B-band magnitude (see Table 4). We get B-V values of 0.71 ± 0.12, 0.69 ± 0.06, 0.87 ± 0.13, and 0.88 ± 0.06 for the four images in order of date. This gives a weighted average of $\langle B-V \rangle = 0.78 \pm 0.04$ mag. This is 1.5-sigma away from the modern value of 0.84 mag. This is fine and is consistent with Pluto remaining the same color since its discovery.

## 4. THE 1953-1955 LIGHT CURVE

Prior to our work in this paper, the earliest reliable light curve for Pluto was that from 1953-1955 (Walker and Hardie 1955). This was made with photoelectric photometry on the McDonald 82" telescope, the 60" and 100" telescopes on Mount Wilson, and the 42" reflector at Lowell Observatory. The resultant 27 V-band points (plus some measured B-V colors) were used to discover the 6.38 day periodicity. The reported magnitudes are only those corrected to mean opposition, which presumably refers to r=39.5 AU, Δ=38.5 AU, and $r\Delta = 1520.7$ AU$^2$. Apparently, no correction was made for the solar phase angle (i.e., the opposition surge). With a linear correction of 0.0372 mag deg$^{-1}$, a best fit sine wave has an average $V_{opp}=14.94$ mag and a peak-to-peak amplitude of 0.12 mag. Their average B-V value is 0.79 mag based on a few measures. With the modern B-V value (0.84 mag), their average $V_{opp}$ will imply that $\langle B_{opp} \rangle = 15.78$ mag in 1953-1955. The uncertainties in their individual magnitudes are roughly 0.02 mag, as seen by the RMS scatter in their magnitudes of Pluto over a few hours (see their Figure 2), while averaged magnitudes will have uncertainties around 0.01 mag with calibration uncertainties dominating.

But how reliable is this early light curve? Robert Hardie was one of the early workers who wrote the definitive analysis procedure (e.g., Hardie 1959) that is essentially used today. Photoelectric photometry of the time was reliable at measuring differential magnitudes with respect to nearby stars of known magnitude. Walker and Hardie measured Pluto's brightness with respect to four identified stars (with their adopted magnitudes quoted). Unfortunately, the measurement of comparison star brightnesses at the time often had systematic errors of up to half a magnitude. For a thorough study of a typical case, Schaefer (1996) examined six independent measures of comparison star sequences for the supernova SN1960F, and found that a thirteenth magnitude star would be reported over a range of 0.69 mag (with an RMS of 0.25 mag). With this, we realize that the Walker and Hardie 1953-1955 light curve might have a large constant offset from the real light curve simply because the comparison stars might have been mismeasured.

To test this, we have made modern CCD measures of the four comparison stars of Walker and Hardie. These observations were made on 18 December 2006 with the 0.9-m SMARTS telescope on Cerro Tololo. A total of 249 observations in B-band and V-band of standard stars (Landolt 1993) were made between airmasses of 1.14 and 1.94 and between B-V colors of -0.29 and 2.19, resulting in an instrumental calibration with RMS scatter of 0.012 and 0.028 mag (in the V-band and B-band) for individual stars and much smaller uncertainties in the calibration relation. The calibration relations were then applied to the four Pluto comparison stars, each of which had been measured six times in the two bands. Our derived magnitudes are essentially identical to those reported by Walker and Hardie, with the average difference being 0.01 mag.

With this, we confirm the comparison star magnitudes used by Walker and Hardie, and we have confidence in the relative photometry from the comparison stars to Pluto. As such, we believe that the reported 1953-1955 light curve is accurate and has no systematic problems.

## 5. PLUTO'S ALBEDO VARIATIONS

As Pluto moves around its orbit, the latitude of the sub-Earth point changes substantially, being closest to the south pole in 1943, crossing the equator around 1987 (with the mutual events), and will be closest to the north pole in 2028. See Figure 3 for full details. In 1933-1934, the latitude of the sub-Earth point on Pluto was -55° to -53° and the Sun-Pluto distance was 40.4 AU. This can be compared to the 1953-1955 light curve with a sub-Earth latitude of-53° and a Sun-Pluto distance of 35.2 AU. That is, from 1933-1934 to 1953-1955, the viewing geometry was identical, while Pluto moved closer to then Sun from 40.4 AU to 35.2 AU. The situation in 1989 (at the last perihelion passage) had a sub-Earth latitude of +4° and a Sun-Pluto distance of 29.6 AU. Currently, the sub-Earth latitude is +40° and the Sun-Pluto distance is 31.5 AU.

Our 1933-1934 light curve pushes the data back by 20 years to a time when Pluto was nearer to aphelion than perihelion. In comparing the 1933-1934 and 1953-1955 light curves, the essentially identical sub-Earth latitudes means that any differences cannot be due to changing viewing geometry. As such, the light curve changes must be due to changes in the albedo, primarily in the southern hemisphere. The only plausible reason for changes in the albedo over this two decade interval is the large change in the Sun-Pluto distance (from 40.4 AU to 35.2 AU).

The first test of albedo changes after 1933 can come from comparing the average $B_{opp}$ between the 1933-1934 and the 1953-1955 light curves. For this, we should use the averages derived from the sine fits, as these avoid biasing due to uneven sampling in phase. The $B_{opp}$ was 15.73±0.01 in 1933-1934 and 15.78±0.01 in 1953-1955. Apparently, with identical sub-Earth latitudes, the surface albedo darkened by roughly 5% in the two decades after 1933.

An alternative test for any change in aldebo would be to compare the observed brightnesses in 1933 with the brightnesses predicted by the modern model of Pluto albedos applied to the geometry of 1933. That is, given the detailed modern maps of Pluto, we can make quantitative and accurate predictions of what the brightness of Pluto should have been in 1933 *if* the surface albedo does not change. The maps were constructed from all the observational material involving light curves, the mutual events in the 1980's, and the maps based on *Hubble Space Telescope* images. The accuracy of the predicted magnitudes for 1933 will be substantially poorer than for modern magnitude predictions as the modern maps do not adequately cover the southern hemisphere which dominates in the 1930's and 1950's. Nevertheless, the modern maps might well be more reliable for comparison with the 1933 light curve than is the 1953-1955 light curve.

We have used our albedo maps of Pluto plus accurate rotation and orbit models (Buie, Young, and Binzel 1997) to predict the $B_{opp}$ values at the times of our early plates. These predictions are presented in Table 3 for each of the nights. The third and fourth columns are the latitude and east longitude of the sub-Earth point ($\lambda_{sub\oplus}$ in degrees). The fifth column presents our model prediction as to the $\langle B_{opp} \rangle$ value. The differences between our observations and our model are given in the last column.

The orbital period of Pluto-Charon is equal to the rotation period of Pluto, and these are 6.387223 ± 0.000017 days (Tholen and Buie 1997). The longitude system is defined relative to the sub-Charon point on Pluto at the time of periapse, with a full description in Buie, Tholen, and Horne (1992). Between a modern epoch (say, the JD 2449000.5 epoch for Charon's orbit, Tholen and Buie, 1997) and our first plate are 21,849.818 days or 3420.86 ± 0.06 rotations. With this, we see that the current ephemeris is of adequate accuracy to know $\lambda_{sub\oplus}$ for the 1933 light curve. The light curves from 1933-1934, 1953-1955, and 1964 all show peak light at near 260° longitude, and this implies that the rotation period is adequately known and that there have been no significant shifts in the bright/dark regions on Pluto.

The amplitude of variation was identical in 1933-1934 and 1953-1955. This is expected due to the identical sub-Earth latitude for the two epochs. The amplitude might have changed had frost sublimation caused asymmetric changes in the aldebo across the southern hemisphere, but this is not seen. The model amplitude for this sub-Earth latitude is 0.12 mag, which is the same as that observed for both epochs.

The model gives the average $B_{opp}$ equal to 15.82 mag. This is 0.09 mag fainter than our observed $\langle B_{opp} \rangle$ for 1933-1934. Similarly, the weighted average $\langle B_{opp} \rangle$-$\langle B_{model} \rangle$ (see the last column of Table 3) is -0.08±0.01. This is to say that the second comparison shows Pluto to have been brighter by 0.08 or 0.09 mag in 1933 than the modern albedo map would have predicted. The uncertainties (both from measurement and systematic errors) appear to be small (at the ~0.01 mag level), so this darkening of Pluto's southern hemisphere appears to be highly significant.

So we now have a picture of Pluto darkening by roughly 5% from 1933-1934 to 1953-1955 and darkening by a further 3-4% to modern times. For the comparison between 1933-1934 and 1953-1955, the geometrical conditions are identical, so the only reason for the 5% darkening can be that Pluto was approaching from 40.4 to 35.2 AU. We know that Pluto at aphelion has a substantial portion of its atmosphere condensed as surface frosts while Pluto at perihelion has these surface frosts sublimed to form a thin atmosphere. Further, we expect that the transient surface frosts will have a higher albedo than the underlying surface material (of indeterminant old age) that has been darkened by interaction with cosmic rays. Thus, we have a coherent picture of Pluto's frost sublimation from 1933 to 1955 with further sublimation until perihelion.

## 6. SEARCH FOR PREDISCOVERY IMAGES

As soon as Pluto was discovered in early 1930, with its orbit being the primary question, it was realized that the best leverage would be to find prediscovery images on old photographs in plate archives around the world. As such, 14 prediscovery images from 1914 to 1927 were identified (Bower 1931). These images are all too poor and too few to hope for any useable photometric information.

In 1982, after substantial experience at using the archival plate collection at Harvard Observatory, Schaefer tried to locate additional prediscovery images. The hope was that images perhaps as early as 1890 might be found so as to substantially increase the known orbital arc for Pluto, with improvements in Pluto's orbit, which had significant residuals at the time. The Harvard collection has half a million plates, and the best from before the date of Pluto's discovery have limits that are deeper than the Palomar Sky Survey. The Harvard collection had been searched in 1930, yet a later search might pull out a missed Pluto image simply because the orbit from the earliest years was known much better in 1982 than in 1930 so a searcher would know better where to look. For the 1982 search, Doug Mink calculated the Pluto ephemeris from 1890 to 1930 at ten day intervals and these positions were plotted on the SAO star atlas for transfer to the plates. A total of 65 plates (from the A, B, I, and MC series) had hopeful limiting magnitudes and covered Pluto's position at the time. Unfortunately, no prediscovery image of Pluto was identified. Despite the deep limits on some Harvard plates, none of the deep plates happened to cover the fields with Pluto and those that did cover Pluto's position did not go deep enough.

In the year 2004, this search of the Harvard collection was repeated with substantially better technique. First, the JPL Horizons ephemeris program provided exact positions for the minute of the middle of each plate exposure. Second, these positions were transferred to the Digital Sky Survey with confidence and high accuracy. With this, we could know exactly where to look (relative to nearby stars) on every plate. Third, Alison Doane had made a complete on-line catalog of all the deepest plates. Fourth, with extensive experience with the Harvard plates (including the discovery of various objects and events missed by previous searchers of these same plates), a more thorough search was possible. Despite these better techniques, no prediscovery image was identified.

One of the claimed prediscovery images (Shapley 1932, Johnstone 1932) is on the Harvard plate MC6858. This plate dates to 1914 November 12.326 taken with a 16" telescope. On the glass side of this plate is a small box drawn in ink and labeled as being of Pluto. The position of this box is accurately placed as with the modern ephemeris, and the location is free from any background stars. On examining the center of the box on the plate, we do not see anything that we would call a stellar image. Rather, the best that it can be described is a "grain enhancement" similar to many in the area. (We have great respect for the utility of getting information from images that are at the plate limiting magnitude. For example, the eclipse period of BT Mon was independently discovered at high significance from images that were mostly very close to the background [Schaefer and Patterson 1983].) Well-developed and well-stored photographic plates do not degrade by any amount, so what we see now is what was seen in 1932. As such, we can only recommend that this observation be removed from the list of prediscovery images and not used for future orbit calculations.

## 7. CONCLUSIONS

We present a reliable B-band light curve for Pluto from 1933-1934 (see Tables 4 and Fig. 2). The average mean-opposition magnitude (with $r\Delta$=1520.7 $AU^2$ and $\phi$=0°) is <$B_{opp}$>=15.73 ± 0.01. The rotational light curve is approximately sinusoidal with a full amplitude of 0.11 ± 0.03 mag. We measure <B-V>=0.78 ± 0.04 mag.

Our light curve is an extension from prior earliest photometric information in 1953-1955 backwards by a total of 20 years. In 1933-1934, Pluto was closer to aphelion than to perihelion, at a time when its surface might be still covered with frosts remaining from its last aphelion passage. From 1933 to 1954, Pluto was fast approaching the Sun with distances of 40.4 to 35.2 AU. The comparison between the 1933-1934 and 1953-1955 light curves (both with the same sub-Sun latitude) should be identical except for any effects of secular changes in Pluto's albedo. Indeed, we see that Pluto has darkened by 5% from 1933-1934 to 1953-1955, followed by a further darkening by 3%-4% until the time of perihelion. We interpret this darkening of Pluto as being caused by the sublimation of relatively bright frosts as Pluto moves from 40.4 AU in to its perihelion.

We report on a remeasurement of the magnitudes for the comparison stars of the 1953-1955 light curve, and we come away with good confidence in the reliability of this light curve as published. We report on two unsuccessful searches of the Harvard plate archive for prediscovery images. We recommend that the 1914 November 12.236 plate be removed from the list of prediscovery images.

## ACKNOWLEDGMENTS

We acknowledge the Carnegie Observatories' photographic and spectral plate archiving project, funded by the Ahmanson Foundation, providing the plates used for this study. We also acknowledge the Harvard College Observatory (with directors Martha Hazen and Alison Doane) for their photographic plates. Our new light curve is only possible by the work of the astronomers from the 1930's and the continuing work at the Carnegie and Harvard plate archives from the 1930's to the present.

We thank Arlo Landolt for his photometry of six of the comparison stars as a check on our magnitudes, for his making available one of the few remaining iris diaphragm photometers, and indeed for his calibration of standard stars vital for our measures of the comparison star magnitudes.

The National Aeronautics and Space Administration provided funds under grants NAG5-13533, NAG5-13369, and NAG5-12835.

TABLE 1

| Plate -Image | UT Date | $UT_{mid}$ | Ap (") | Exp (min) | Comments | $\lambda_{sub\oplus}$ | r (AU) | Δ (AU) | α (°) | Δm (mag) |
|---|---|---|---|---|---|---|---|---|---|---|
| B145 | 1933 Mar 19 | 04:22 | 60" | 10 | good | 90 | 40.56 | 40.16 | 1.29 | -0.20 |
| B147 | 1933 Mar 19 | 05:37 | 60" | 10 | ok | 87 | 40.56 | 40.16 | 1.29 | -0.20 |
| B148 | 1933 Mar 19 | 06:36 | 60" | 10 | poor | 85 | 40.56 | 40.16 | 1.29 | -0.20 |
| B149 | 1933 Mar 19 | 07:21 | 60" | 10 | poor | 83 | 40.56 | 40.16 | 1.29 | -0.20 |
| B150 | 1933 Mar 19 | 07:52 | 60" | 10 | very poor | 82 | 40.56 | 40.16 | 1.29 | -0.20 |
| B151 | 1933 Mar 20 | 03:51 | 60" | 5 | good | 35 | 40.56 | 40.18 | 1.30 | -0.20 |
| B152-A | 1933 Mar 20 | 04:10 | 60" | 10 | ok | 34 | 40.56 | 40.18 | 1.30 | -0.20 |
| B152-B | 1933 Mar 20 | 04:22 | 60" | 10 | ok | 34 | 40.56 | 40.18 | 1.30 | -0.20 |
| B153 | 1933 Mar 20 | 05:26 | 60" | 10 | good | 32 | 40.56 | 40.18 | 1.30 | -0.20 |
| B284B | 1933 Oct 14 | 11:51 | 100" | 11 | great | 168 | 40.43 | 40.48 | 1.41 | -0.22 |
| B285B | 1933 Oct 14 | 11:56 | 100" | 2 | good | 168 | 40.43 | 40.48 | 1.41 | -0.22 |
| B286B | 1933 Oct 14 | 12:39 | 100" | 2 | good | 167 | 40.43 | 40.48 | 1.41 | -0.22 |
| P2=a | 1933 Oct 26 | 12:27 | 60" | 10 | good | 210 | 40.42 | 40.27 | 1.40 | -0.20 |
| P1=b | 1933 Oct 26 | 12:39 | 60" | 10 | good | 210 | 40.42 | 40.27 | 1.40 | -0.20 |
| P4=c | 1933 Oct 27 | 12:21 | 60" | 10 | good | 154 | 40.42 | 40.25 | 1.39 | -0.20 |
| P3=d | 1933 Oct 27 | 12:33 | 60" | 10 | good | 154 | 40.42 | 40.25 | 1.39 | -0.20 |
| B306B | 1933 Nov 16 | 12:10 | 100" | 3 | good | 108 | 40.40 | 39.91 | 1.23 | -0.18 |
| B305B | 1933 Nov 16 | 12:29 | 100" | 3 | good | 107 | 40.40 | 39.91 | 1.23 | -0.18 |
| B290-D | 1933 Nov 18 | 09:20 | 60" | 4 | good | 1 | 40.40 | 39.88 | 1.21 | -0.17 |
| B290-C | 1933 Nov 18 | 09:27 | 60" | 6.75 | good | 1 | 40.40 | 39.88 | 1.21 | -0.17 |
| B290-B | 1933 Nov 18 | 09:35 | 60" | 11.5 | good | 1 | 40.40 | 39.88 | 1.21 | -0.17 |
| B290-A | 1933 Nov 18 | 10:23 | 60" | 56 | V-band | 359 | 40.40 | 39.88 | 1.21 | -0.17 |
| B291-D | 1933 Nov 18 | 11:10 | 60" | 4 | good | 357 | 40.40 | 39.88 | 1.21 | -0.17 |
| B291-C | 1933 Nov 18 | 11:17 | 60" | 6.75 | good | 357 | 40.40 | 39.88 | 1.21 | -0.17 |
| B291-B | 1933 Nov 18 | 11:25 | 60" | 11.5 | good | 357 | 40.40 | 39.88 | 1.21 | -0.17 |
| B291-A | 1933 Nov 18 | 12:19 | 60" | 56 | V-band | 354 | 40.40 | 39.88 | 1.21 | -0.17 |
| B294-A | 1933 Nov 19 | 09:48 | 60" | 4 | good | 304 | 40.40 | 39.87 | 1.19 | -0.17 |
| B294-B | 1933 Nov 19 | 09:55 | 60" | 6.75 | good | 304 | 40.40 | 39.87 | 1.19 | -0.17 |
| B294-C | 1933 Nov 19 | 10:30 | 60" | 56 | V-band | 302 | 40.40 | 39.87 | 1.19 | -0.17 |
| B295-A | 1933 Nov 19 | 11:22 | 60" | 4 | good | 300 | 40.40 | 39.87 | 1.19 | -0.17 |
| B295-B | 1933 Nov 19 | 11:29 | 60" | 6.75 | good | 300 | 40.40 | 39.87 | 1.19 | -0.17 |
| B295-C | 1933 Nov 19 | 12:05 | 60" | 56 | V-band | 300 | 40.40 | 39.87 | 1.19 | -0.17 |
| B296 | 1933 Nov 19 | 12:57 | 60" | 10 | good | 297 | 40.40 | 39.87 | 1.19 | -0.17 |
| B297 | 1933 Nov 19 | 13:10 | 60" | 5 | good | 268 | 40.40 | 39.87 | 1.19 | -0.17 |
| M2063 | 1933 Nov 23 | 07:35 | … | 10 | good | 84 | 40.40 | 39.81 | 1.14 | -0.17 |
| M2068 | 1933 Nov 24 | 09:43 | … | 60 | var. bck | 22 | 40.40 | 39.80 | 1.12 | -0.17 |
| MA3429 | 1934 Jan 18 | 04:49 | 12" | 67 | good | 175 | 40.36 | 39.38 | 0.09 | -0.10 |
| MC27075 | 1934 Feb 7 | 03:06 | 16" | 30 | great | 133 | 40.35 | 39.45 | 0.58 | -0.12 |
| MC27077 | 1934 Feb 7 | 04:19 | 16" | 30 | great | 130 | 40.35 | 39.45 | 0.58 | -0.12 |
| MC27095 | 1934 Feb 18 | 03:14 | 16" | 30 | great | 233 | 40.34 | 39.53 | 0.82 | -0.14 |
| MC27097 | 1934 Feb 18 | 04:38 | 16" | 30 | great | 229 | 40.34 | 39.53 | 0.82 | -0.14 |
| MA3601 | 1934 Mar 8 | 01:49 | 12" | 60 | good | 302 | 40.33 | 39.74 | 1.14 | -0.16 |
| MC27120 | 1934 Mar 10 | 02:14 | 16" | 30 | great | 188 | 40.33 | 39.77 | 1.17 | -0.16 |

TABLE 2

Comparison star magnitudes

| Star | RA2000 | Dec2000 | B | V |
|---|---|---|---|---|
| 1-01 | 7 36 40.96 | +22 35 40.6 | 14.03 | 13.52 |
| 1-02 | 7 36 43.09 | +22 34 16.7 | 16.68 | 15.87 |
| 1-03 | 7 36 40.32 | +22 33 46.2 | 14.57 | 14.11 |
| 1-04 | 7 36 36.74 | +22 34 23.3 | 16.72 | 16.12 |
| 1-05 | 7 36 31.99 | +22 34 53.0 | 15.45 | 14.70 |
| 1-06 | 7 36 37.66 | +22 36 07.4 | 15.32 | 14.79 |
| 1-07 | 7 36 40.44 | +22 36 34.7 | 15.86 | 15.09 |
| 1-08 | 7 36 37.02 | +22 37 55.9 | 16.11 | 15.03 |
| 1-09 | 7 36 36.98 | +22 38 29.0 | 16.25 | 15.35 |
| 1-10 | 7 36 50.88 | +22 34 51.6 | 17.11 | 16.02 |
| 1-11 | 7 36 53.99 | +22 33 58.6 | 16.23 | 15.34 |
| 1-12 | 7 36 42.70 | +22 31 56.9 | 16.11 | 15.30 |
| 1-13 | 7 36 45.65 | +22 32 29.2 | 16.44 | 15.97 |
| 1-14 | 7 36 29.03 | +22 34 37.5 | 16.02 | 15.43 |
| 1-15 | 7 36 33.43 | +22 32 48.2 | 16.13 | 15.56 |
| 1-16 | 7 36 31.39 | +22 32 03.0 | 16.50 | 15.70 |
| 1-17 | 7 36 50.69 | +22 32 49.4 | 16.98 | 16.22 |
| 2-01 | 7 51 11.78 | +22 15 21.6 | 18.14 | 17.02 |
| 2-02 | 7 51 15.45 | +22 15 31.7 | 17.44 | 16.65 |
| 2-03 | 7 51 19.66 | +22 15 49.4 | 17.79 | 17.07 |
| 2-04 | 7 51 10.75 | +22 16 03.1 | 16.12 | 15.51 |
| 2-05 | 7 51 10.60 | +22 16 29.3 | 16.07 | 15.42 |
| 2-06 | 7 51 04.87 | +22 16 54.7 | 15.26 | 14.70 |
| 2-07 | 7 51 04.25 | +22 17 32.9 | 16.34 | 15.78 |
| 2-08 | 7 51 19.85 | +22 12 03.1 | 16.36 | 15.36 |
| 2-09 | 7 51 20.42 | +22 10 54.2 | 17.05 | 16.36 |
| 2-10 | 7 51 15.08 | +22 13 43.8 | 13.93 | 13.30 |
| 2-11 | 7 51 49.00 | +22 12 21.0 | 16.41 | 14.86 |
| 2-12 | 7 51 49.70 | +22 11 07.9 | 16.55 | 15.83 |
| 2-13 | 7 51 36.24 | +22 08 23.6 | 16.04 | 15.48 |
| 2-14 | 7 51 17.06 | +22 14 02.9 | 16.37 | 14.87 |
| 2-15 | 7 51 32.63 | +22 15 58.3 | 15.62 | 14.46 |
| 2-18 | 7 51 22.25 | +22 17 01.4 | 15.54 | 13.78 |
| 2-19 | 7 51 20.53 | +22 17 46.9 | 16.49 | 15.81 |
| 2-20 | 7 51 34.94 | +22 17 32.1 | 15.15 | 14.41 |
| 2-21 | 7 51 34.05 | +22 18 23.2 | 15.42 | 14.77 |
| 2-A | 7 51 31.28 | +22 12 48.1 | 14.75 | 14.14 |
| 2-B | 7 51 35.37 | +22 10 52.9 | 14.78 | 14.00 |
| 2-C | 7 51 34.08 | +22 09 32.9 | 15.60 | 14.77 |
| 2-D | 7 51 30.74 | +22 11 14.6 | 15.75 | 15.08 |
| 2-E | 7 51 36.06 | +22 09 47.9 | 16.20 | 15.51 |
| 2-F | 7 51 46.47 | +22 11 23.0 | 15.69 | 14.84 |
| 3-02 | 7 46 33.25 | +22 35 25.5 | 13.45 | 12.72 |
| 3-03 | 7 46 43.17 | +22 34 40.2 | 16.56 | 16.06 |
| 3-04 | 7 46 37.40 | +22 36 13.7 | 15.18 | 14.55 |
| 3-05 | 7 46 32.74 | +22 36 38.7 | 15.09 | 14.18 |
| 3-06 | 7 46 45.88 | +22 35 40.2 | 15.20 | 14.74 |
| 3-07 | 7 46 48.61 | +22 31 32.6 | 14.94 | 14.33 |
| 3-08 | 7 46 50.83 | +22 30 52.6 | 16.35 | 15.73 |
| 3-09 | 7 46 54.66 | +22 29 40.6 | 14.52 | 14.06 |
| 3-10 | 7 46 50.70 | +22 29 40.0 | 16.54 | 15.92 |
| 3-11 | 7 46 31.43 | +22 28 04.3 | 12.86 | 12.44 |
| 3-12 | 7 46 55.58 | +22 34 41.1 | 13.96 | 13.43 |
| 3-13 | 7 46 20.16 | +22 34 55.6 | 16.91 | 16.21 |
| 3-14 | 7 46 27.9 | +22 31 04 | 16.27 | 15.57 |
| 3-15 | 7 46 57.26 | +22 34 04.5 | 14.73 | 14.03 |
| 3-16 | 7 46 55.25 | +22 34 00.4 | 16.54 | 15.90 |
| 3-19 | 7 46 28.94 | +22 32 02.1 | 16.51 | 15.82 |
| 3-20 | 7 46 40.82 | +22 36 06.5 | 17.31 | 16.33 |
| 4-01 | 7 44 51.93 | +22 39 34.4 | 14.31 | 13.60 |
| 4-02 | 7 44 55.30 | +22 37 18.5 | 14.86 | 14.09 |
| 4-03 | 7 44 49.58 | +22 37 20.2 | 15.81 | 15.02 |
| 4-04 | 7 44 59.89 | +22 36 43.4 | 14.08 | 13.49 |
| 4-05 | 7 44 49.49 | +22 41 09.0 | 15.71 | 15.04 |
| 4-06 | 7 44 58.45 | +22 41 39.9 | 16.05 | 15.55 |
| 4-07 | 7 44 51.14 | +22 41 47.2 | 15.12 | 14.20 |
| 4-09 | 7 45 04.42 | +22 38 30.4 | 16.24 | 15.66 |
| 4-10 | 7 45 00.00 | +22 39 23.1 | 17.39 | 16.74 |
| 4-11 | 7 45 10.48 | +22 38 55.6 | 16.41 | 15.69 |
| 4-12 | 7 44 49.73 | +22 36 19.1 | 15.78 | 15.05 |
| 4-14 | 7 44 43.02 | +22 40 39.6 | 17.61 | 16.31 |
| 4-15 | 7 44 36.15 | +22 40 53.7 | 17.20 | 16.27 |
| 4-16 | 7 44 46.58 | +22 35 04.1 | 17.28 | 16.64 |
| 4-17 | 7 45 09.28 | +22 37 25.5 | 16.80 | 16.07 |
| 4-18 | 7 45 08.88 | +22 39 57.0 | 18.09 | 17.13 |
| 4-19 | 7 44 38.06 | +22 36 22.7 | 13.14 | 12.70 |
| 4-20 | 7 44 45.09 | +22 39 31.9 | 17.53 | 16.89 |
| 5-01 | 7 44 01.54 | +22 43 07.5 | 14.24 | 13.70 |
| 5-02 | 7 43 59.97 | +22 42 36.5 | 16.92 | 16.14 |
| 5-03 | 7 44 02.41 | +22 40 34.1 | 16.07 | 15.47 |
| 5-04 | 7 44 09.14 | +22 39 33.7 | 17.21 | 16.66 |
| 5-05 | 7 44 16.86 | +22 42 16.5 | 13.15 | 12.44 |
| 5-06 | 7 44 26.04 | +22 43 10.6 | 13.81 | 12.98 |
| 5-07 | 7 43 48.36 | +22 42 09.3 | 14.03 | 13.51 |
| 5-08 | 7 43 42.85 | +22 41 29.7 | 12.04 | 11.67 |
| 5-09 | 7 43 46.66 | +22 43 14.0 | 17.60 | 16.45 |
| 5-10 | 7 43 43.16 | +22 42 54.9 | 17.05 | 16.43 |
| 5-11 | 7 43 51.20 | +22 39 06.2 | 14.76 | 14.20 |
| 5-12 | 7 43 51.55 | +22 37 18.4 | 13.57 | 13.09 |
| 5-13 | 7 43 54.86 | +22 36 01.6 | 17.16 | 16.55 |
| 5-15 | 7 44 28.93 | +22 36 10.5 | 13.47 | 12.71 |
| 5-16 | 7 44 01.18 | +22 38 01.2 | 14.64 | 14.13 |
| 5-17 | 7 44 00.16 | +22 39 06.5 | 16.58 | 15.89 |
| 5-18 | 7 44 02.63 | +22 36 30.0 | 17.19 | 16.36 |
| 5-19 | 7 44 03.47 | +22 37 17.7 | 13.78 | 13.26 |
| 5-20 | 7 44 02.06 | +22 35 58.5 | 17.36 | 16.43 |
| 6-01 | 7 42 57.45 | +22 46 31.0 | 15.38 | 14.84 |
| 6-02 | 7 42 53.29 | +22 46 32.1 | 16.36 | 15.73 |
| 6-03 | 7 42 50.26 | +22 46 45.8 | 15.79 | 15.07 |
| 6-04 | 7 42 54.39 | +22 44 50.2 | 15.06 | 14.61 |
| 6-05 | 7 42 50.24 | +22 43 24.6 | 16.84 | 15.48 |
| 6-06 | 7 42 54.81 | +22 43 36.3 | 16.28 | 15.69 |
| 6-07 | 7 42 59.13 | +22 47 33.6 | 13.57 | 12.31 |
| 6-08 | 7 43 00.12 | +22 48 21.2 | 15.29 | 14.46 |
| 6-09 | 7 42 43.14 | +22 44 38.3 | 15.00 | 14.49 |
| 6-10 | 7 43 10.09 | +22 43 54.7 | 14.74 | 14.22 |
| 6-11 | 7 43 01.61 | +22 41 33.7 | 14.85 | 13.82 |
| 6-12 | 7 43 14.80 | +22 46 58.0 | 13.69 | 13.13 |
| 6-14 | 7 42 51.70 | +22 40 43.2 | 14.68 | 13.75 |
| 6-15 | 7 43 09.74 | +22 48 57.0 | 15.00 | 14.53 |
| 6-17 | 7 43 18.92 | +22 45 16.9 | 16.25 | 15.69 |
| 6-18 | 7 42 55.89 | +22 40 34.6 | 16.98 | 16.29 |
| 6-19 | 7 42 44.32 | +22 46 00.4 | 17.80 | 16.81 |
| 6-20 | 7 42 49.30 | +22 48 23.4 | 17.41 | 16.79 |

TABLE 3

| Plate-Image | Band | $JD_{mid}$ | Observed m (mag) | $m_{opp}$ (mag) |
|---|---|---|---|---|
| B145 | B | 2427150.682 | 15.96 ± 0.03 | 15.76 |
| B147 | B | 2427150.734 | 16.05 ± 0.15 | 15.85 |
| B148 | B | 2427150.775 | 16.02 ± 0.06 | 15.82 |
| B149 | B | 2427150.806 | 16.02 ± 0.05 | 15.82 |
| B150 | B | 2427150.828 | 16.11 ± 0.07 | 15.91 |
| B151 | B | 2427151.660 | 15.89 ± 0.06 | 15.69 |
| B152-A | B | 2427151.674 | 16.03 ± 0.07 | 15.83 |
| B152-B | B | 2427151.682 | 15.97 ± 0.08 | 15.77 |
| B153 | B | 2427151.726 | 15.87 ± 0.06 | 15.66 |
| B284B | B | 2427359.994 | 15.86 ± 0.06 | 15.64 |
| B285B | B | 2427359.997 | 15.81 ± 0.08 | 15.59 |
| B286B | B | 2427360.027 | 15.87 ± 0.09 | 15.65 |
| P2=a | B | 2427372.019 | 15.95 ± 0.18 | 15.75 |
| P1=b | B | 2427372.027 | 15.79 ± 0.08 | 15.58 |
| P4=c | B | 2427373.015 | 15.92 ± 0.15 | 15.72 |
| P3=d | B | 2427373.023 | 15.75 ± 0.10 | 15.55 |
| B306B | B | 2427393.007 | 15.84 ± 0.09 | 15.67 |
| B305B | B | 2427393.020 | 15.92 ± 0.10 | 15.74 |
| B290-D | B | 2427394.889 | 15.93 ± 0.04 | 15.75 |
| B290-C | B | 2427394.894 | 15.95 ± 0.09 | 15.78 |
| B290-B | B | 2427394.899 | 15.89 ± 0.06 | 15.71 |
| B290-A | V | 2427394.933 | 15.19 ± 0.12 | 15.02 |
| B291-D | B | 2427394.965 | 15.93 ± 0.06 | 15.75 |
| B291-C | B | 2427394.970 | 15.88 ± 0.05 | 15.70 |
| B291-B | B | 2427394.976 | 15.87 ± 0.06 | 15.69 |
| B291-A | V | 2427395.013 | 15.21 ± 0.06 | 15.04 |
| B294-A | B | 2427395.908 | 15.91 ± 0.07 | 15.73 |
| B294-B | B | 2427395.913 | 15.85 ± 0.09 | 15.68 |
| B294-C | V | 2427395.938 | 15.00 ± 0.13 | 14.83 |
| B295-A | B | 2427395.974 | 15.84 ± 0.04 | 15.67 |
| B295-B | B | 2427395.978 | 15.94 ± 0.07 | 15.77 |
| B295-C | V | 2427396.003 | 14.99 ± 0.06 | 14.82 |
| B296 | B | 2427396.040 | 15.80 ± 0.09 | 15.63 |
| B297 | B | 2427396.549 | 15.91 ± 0.17 | 15.74 |
| M2063 | B | 2427399.816 | 16.08 ± 0.08 | 15.92 |
| M2068 | B | 2427400.905 | 15.60 ± 0.04 | 15.44 |
| MA3429 | B | 2427455.701 | 15.86 ± 0.15 | 15.75 |
| MC27075 | B | 2427475.629 | 15.96 ± 0.10 | 15.84 |
| MC27077 | B | 2427475.680 | 16.06 ± 0.10 | 15.93 |
| MC27095 | B | 2427486.635 | 15.97 ± 0.06 | 15.83 |
| MC27097 | B | 2427486.693 | 15.87 ± 0.09 | 15.74 |
| MA3601 | B | 2427504.576 | 15.66 ± 0.11 | 15.50 |
| MC27120 | B | 2427506.593 | 15.90 ± 0.04 | 15.74 |

TABLE 4

Nightly averaged B-band magnitudes; observed and modeled

| $\langle JD_{mid} \rangle$ | $\langle B_{opp} \rangle$ | $\phi_{sub\oplus}$ | $\lambda_{sub\oplus}$ | $\langle B_{model} \rangle$ | $\langle B_{model} \rangle - \langle B_{opp} \rangle$ |
|---|---|---|---|---|---|
| 2427150.77 | 15.80 ± 0.02 | -53 | 86 | 15.88 | -0.08 ± 0.02 |
| 2427151.69 | 15.73 ± 0.03 | -53 | 34 | 15.83 | -0.10 ± 0.03 |
| 2427360.01 | 15.63 ± 0.04 | -55 | 168 | 15.81 | -0.18 ± 0.04 |
| 2427372.02 | 15.61 ± 0.07 | -55 | 210 | 15.78 | -0.17 ± 0.07 |
| 2427373.02 | 15.60 ± 0.08 | -55 | 154 | 15.83 | -0.23 ± 0.08 |
| 2427393.01 | 15.70 ± 0.07 | -55 | 107 | 15.87 | -0.17 ± 0.07 |
| 2427394.93 | 15.73 ± 0.02 | -55 | 359 | 15.79 | -0.06 ± 0.02 |
| 2427395.96 | 15.69 ± 0.03 | -55 | 301 | 15.79 | -0.10 ± 0.03 |
| 2427396.55 | 15.74 ± 0.17 | -55 | 268 | 15.78 | -0.04 ± 0.17 |
| 2427399.82 | 15.92 ± 0.08 | -55 | 84 | 15.86 | 0.06 ± 0.08 |
| 2427455.70 | 15.75 ± 0.15 | -55 | 175 | 15.76 | -0.01 ± 0.15 |
| 2427475.65 | 15.88 ± 0.07 | -54 | 131 | 15.84 | 0.04 ± 0.07 |
| 2427486.66 | 15.80 ± 0.05 | -54 | 231 | 15.76 | 0.04 ± 0.05 |
| 2427504.58 | 15.50 ± 0.11 | -54 | 302 | 15.79 | -0.29 ± 0.11 |
| 2427506.59 | 15.74 ± 0.04 | -54 | 188 | 15.79 | -0.05 ± 0.04 |

TABLE 5

Phase averaged B-band magnitudes

| $\langle \lambda_{sub\oplus} \rangle$ | $\langle B_{opp} \rangle$ |
|---|---|
| 34 | 15.73 ± 0.03 |
| 85 | 15.81 ± 0.02 |
| 119 | 15.79 ± 0.05 |
| 175 | 15.68 ± 0.03 |
| 224 | 15.74 ± 0.04 |
| 300 | 15.68 ± 0.03 |
| 359 | 15.73 ± 0.02 |

**Figure 1.** Calibration curves for three Pluto plates.
The calibration curves are plots of $R^2$-$R_b^2$ versus magnitude, with this being linear over the range of interest. The three panels display the calibration curves for our best case, a typical plate, and our worst case. The measurement errors for both the IDP readings and the comparison star magnitudes are much smaller than the dots, with the observed scatter being entirely from random grain noise in the emulsion. A straight line is fit within a range around that of Pluto, and this fit is displayed as a diagonal straight line segment in each figure. The vertical line indicates the observed $R^2$-$R_b^2$ value for Pluto, and this intersects the best fit calibration curve so as to define the height of the horizontal line which indicates the derived magnitude for Pluto. In the top panel, the best case plate is B145 with $R^2$-$R_b^2$=56.5±0.5 for Pluto and hence B=15.96±0.03 (as indicated by the vertical and horizontal lines). In the middle panel, the typical case plate is B152-A with $R^2$-$R_b^2$=40.4±0.2 for Pluto and hence B=16.03±0.07. In the bottom panel, the worst case plate is P2 with $R^2$-$R_b^2$=33.0±0.2 for Pluto and hence B=15.95±0.18.

**Figure 2.** Rotational light curve for Pluto averaged over phase.
Our nightly average $B_{opp}$ values can be averaged together for nights with similar phase (see Table 5). The plot shows two rotational periods (of 6.38 days each) with the magnitudes from Table 5 plotted twice. The central region, for longitudes 0°-360°, are represented by filled diamonds for the data and solid curves for the best fit model. The best fit sinewave is superposed, with a peak-to-peak amplitude of 0.11 ± 0.03 mag. Note that the reduced chi-square of this empirical model is near unity, and this implies that our quoted error bars are reliable.

**Figure 3.** Viewing geometry for Pluto from 1920-2040.
In the late 1980's, Pluto passed through perihelion and had the mutual events between Pluto and Charon. As Pluto orbits the Sun, our Earthly viewpoint shifts. In the late 1980's the sub-Earth latitude on Pluto was near Pluto's equator, while since then we have been primarily looking at Pluto's northern hemisphere. The sub-Earth latitude was identical for 1933-1934 and 1953-1955, so any changes in the light curve cannot be due to changing viewing geometry. As such, any changes in brightness (corrected to mean opposition) must come from a real change in the albedo of Pluto's surface. Our observations demonstrate that Pluto's albedo darkened by 5% in this two decade interval, and this change can only be due to the changing Pluto-Sun distance (falling from 40.4 to 35.2 AU). We interpret this as being caused by the relatively bright surface frosts sublimating as Pluto's atmosphere forms with the approach of perihelion.

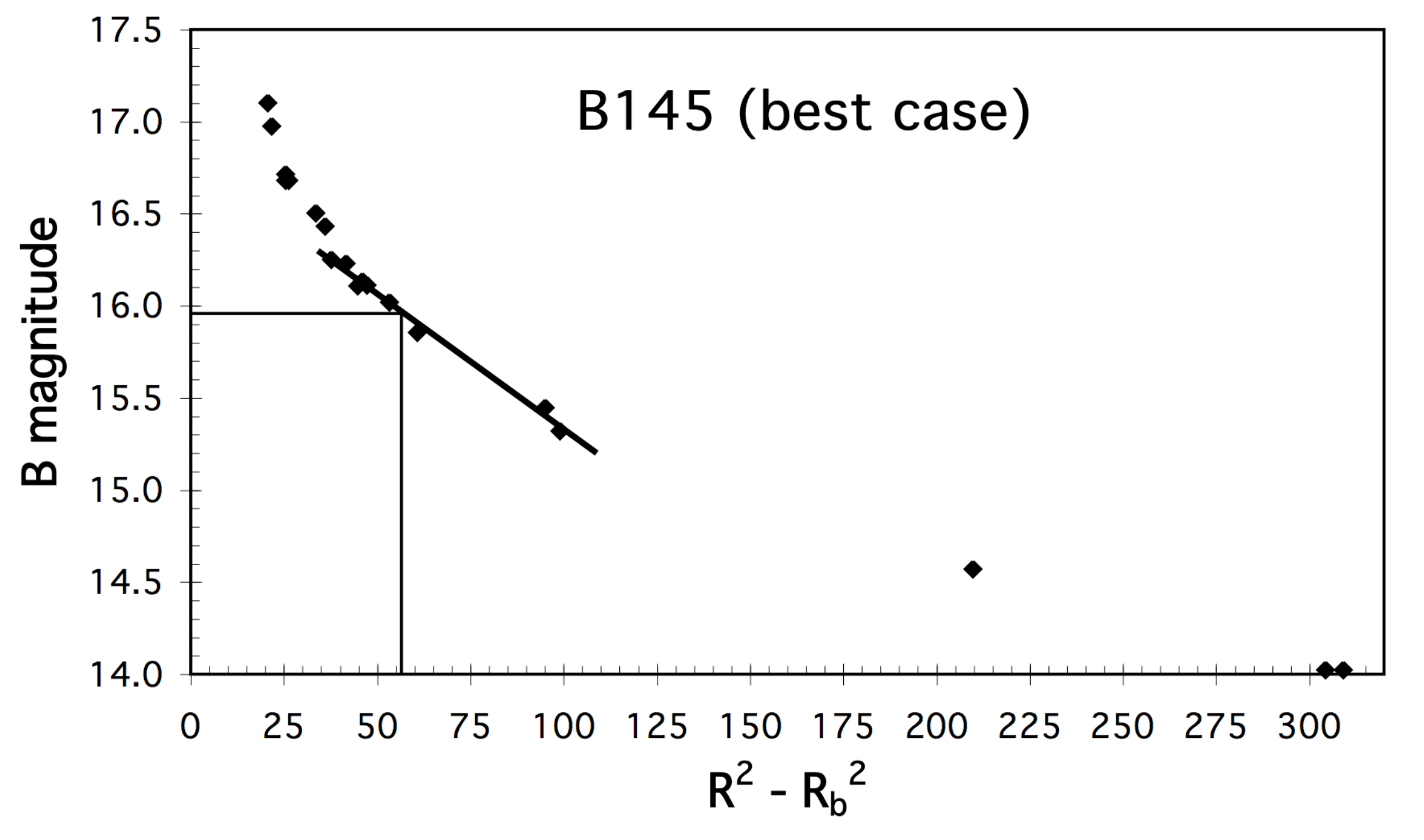

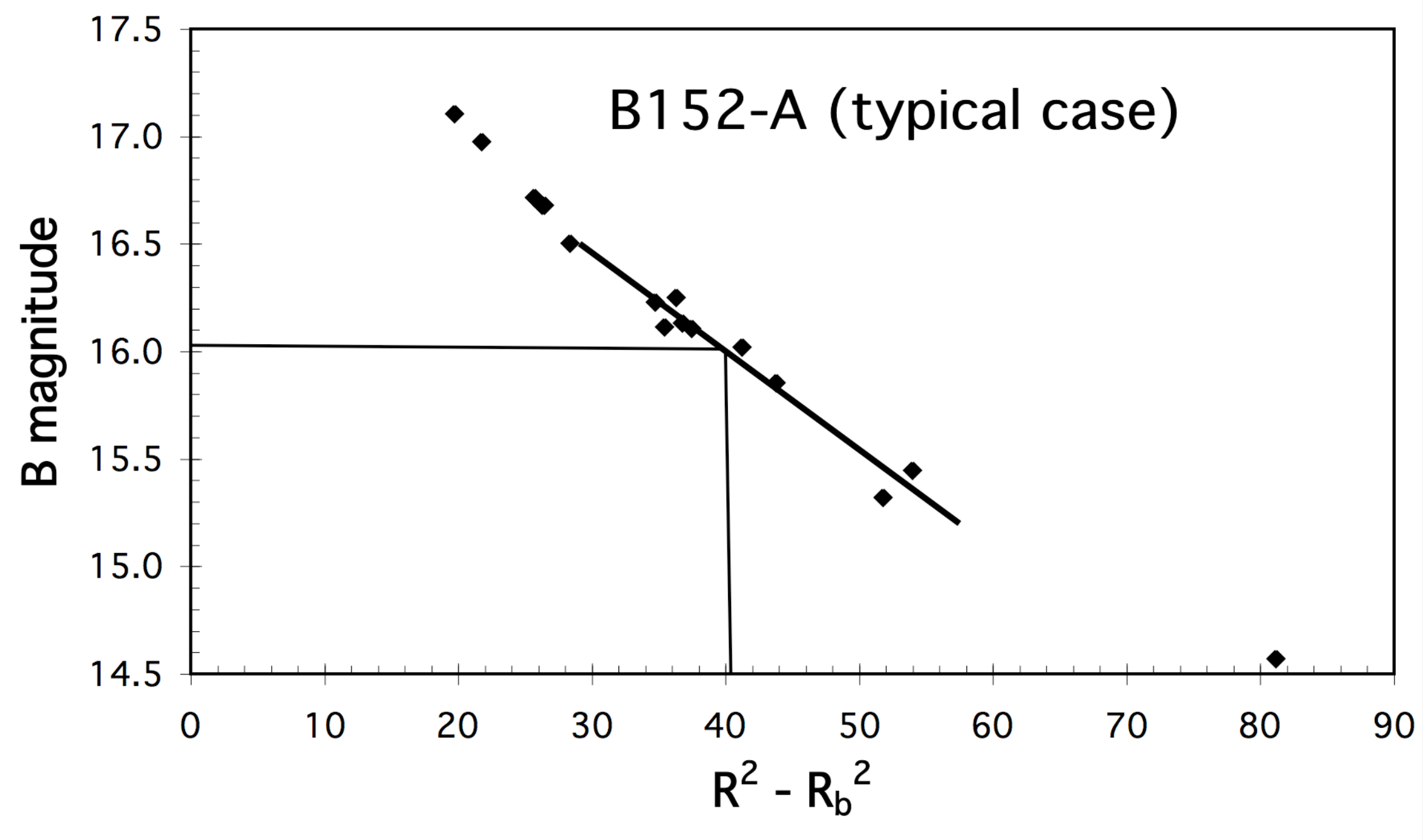

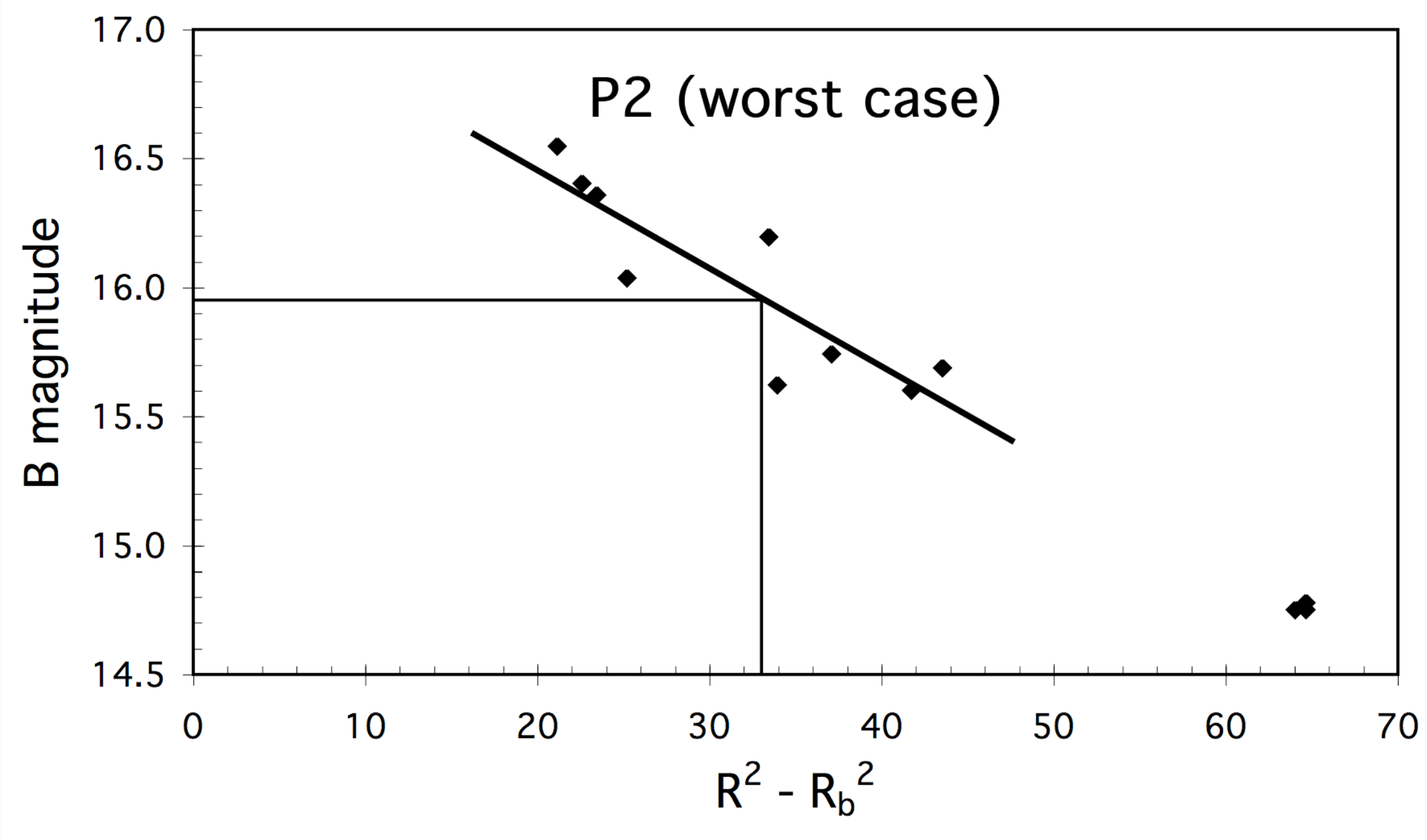

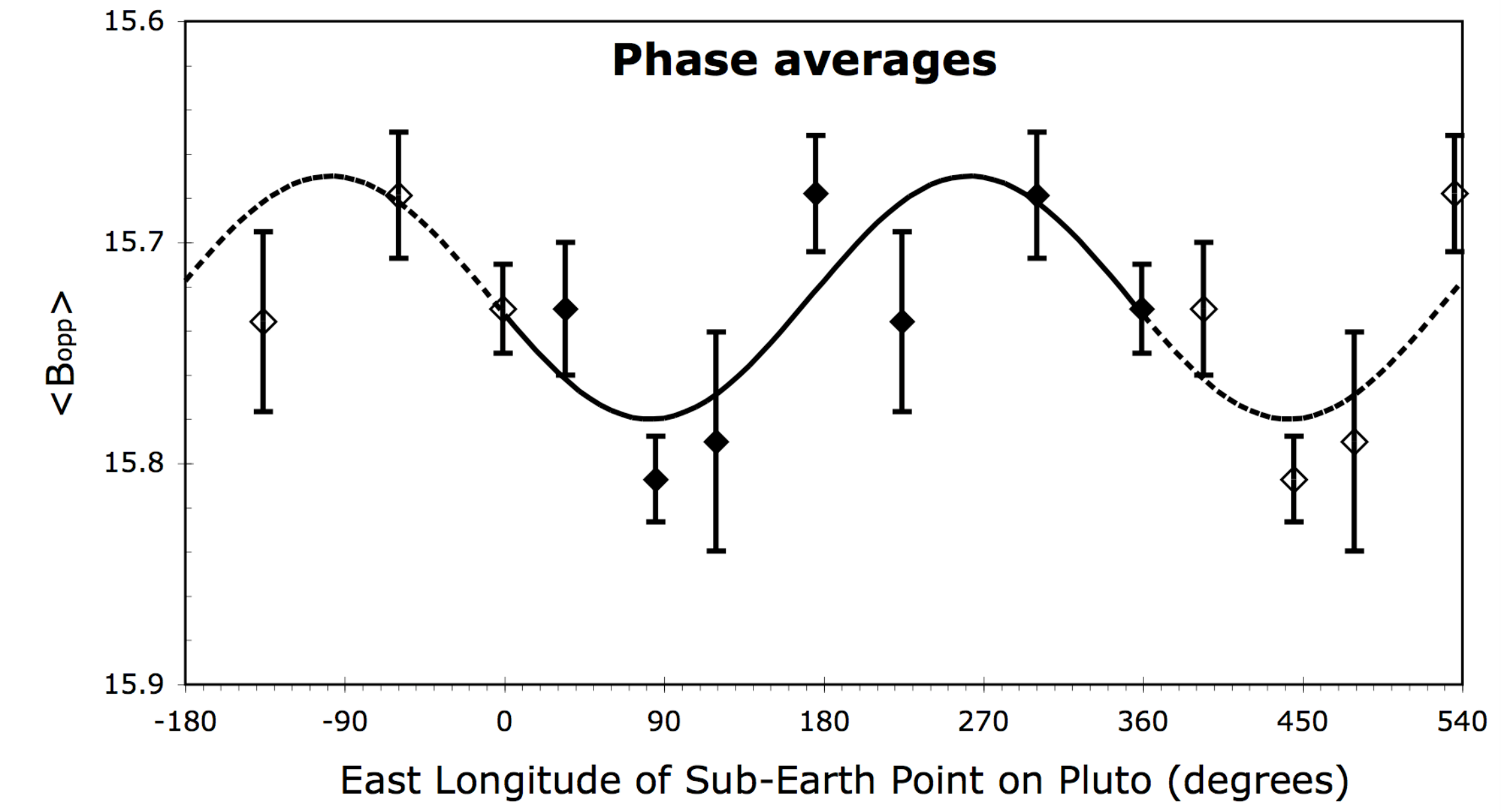
Phase averages
<$B_{opp}$>
15.6
15.7
15.8
15.9
-180
-90
0
90
180
270
360
450
540
East Longitude of Sub-Earth Point on Pluto (degrees)

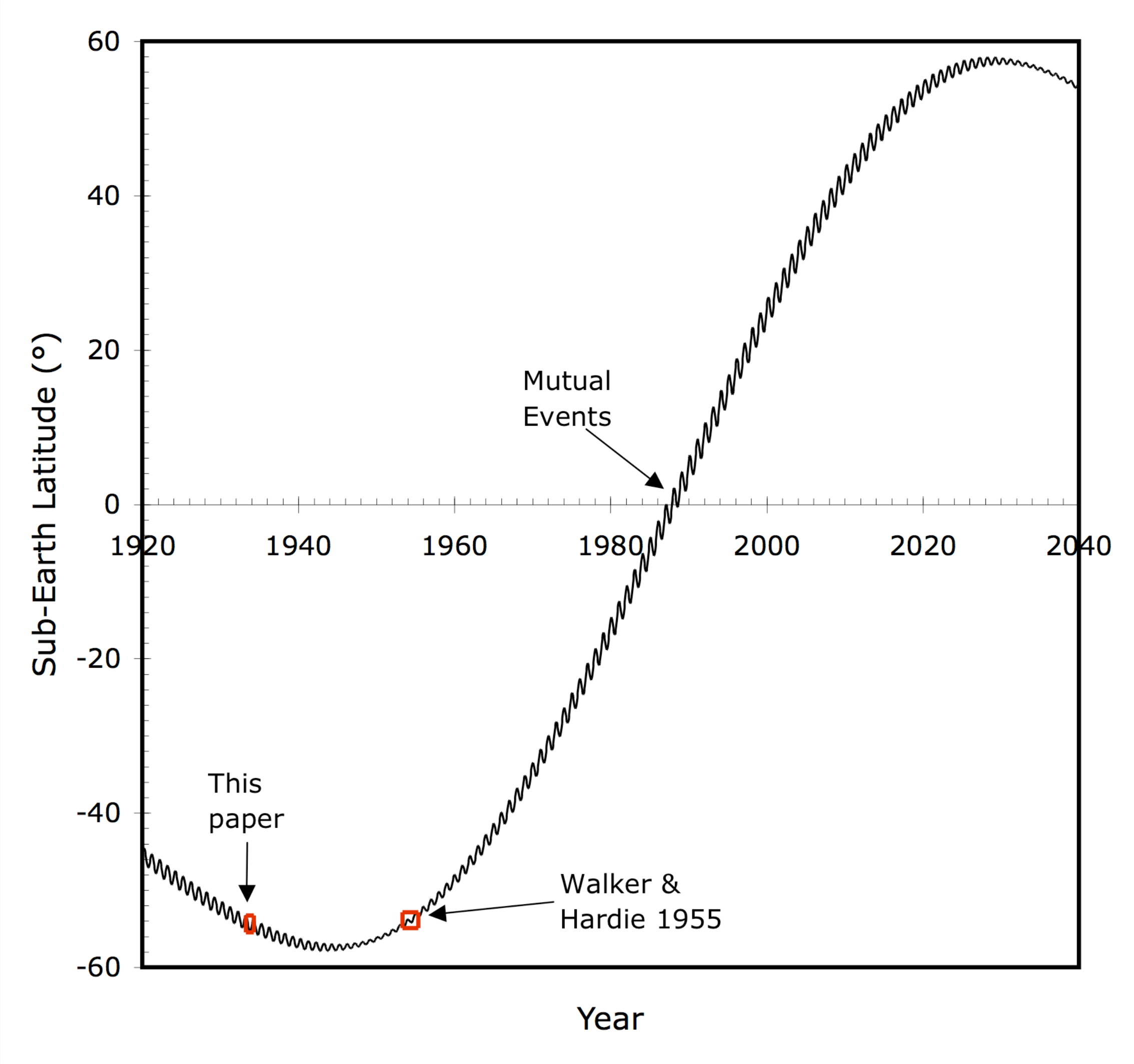